\documentclass[12pt]{article}
\newcommand{\half}{\frac{1}{2}}

\newcommand{\BV}{\left(\begin{array}{c}}
\newcommand{\EV}{\end{array}\right)}
\newcommand{\BM}{\left(\begin{array}{cc}}

\newcommand{\beq}{\begin{equation}}               
\newcommand{\eeq}{\end{equation}}                 
\newcommand{\bqry}{\begin{eqnarray}}              
\newcommand{\eqry}{\end{eqnarray}}                
\newcommand{\bqryn}{\begin{eqnarray*}}            
\newcommand{\eqryn}{\end{eqnarray*}}              
\begin{document}
\begin{titlepage}

\title{\bf Implications of Quark-Lepton Symmetry for \\
Neutrino Masses and Oscillations}
\author{
T. Goldman\thanks{\small \em E-mail:
t.goldman@post.harvard.edu}\\{\small \em 
MS B283, Theoretical Division, Los Alamos National Laboratory}\\ 
{\small \em Los Alamos, New Mexico 87545 USA}\\
G.J. Stephenson, Jr.\thanks{\small \em E-mail:
GJS@baryon.phys.unm.edu}\\
{\small \em Dept. of Physics \& Astronomy, University of New Mexico}\\ 
{\small \em Albuquerque, NM 87131 USA}\\
B.H.J. McKellar\thanks{\small \em E-mail:
mckellar@physics.unimelb.EDU.AU}\\
{\small \em School of Physics, University of Melbourne}\\ 
{\small \em Parkville, Victoria 3052 Australia}
}

\date{February 22, 2000}
\maketitle

\vspace{-5.5in}
\flushright{LA-UR-00-908}
\vspace{-0.25in}
\flushright{UM-P-2000/007}
\vspace{-0.25in}
\flushright{nucl-th/0002053}
\vspace*{4.2in}

\begin{abstract}

We identify a plausible scenario based on quark-lepton symmetry which
correlates long baseline oscillations with maximal mixing to sterile
neutrinos. The implication for the Sudbury Neutrino Observatory (SNO)
is that the neutral current signal will be found to suffer the same
suppression from the Standard Solar Model prediction as obtains for the
charged current signal. Flavor mixing among active neutrinos is
expected to occur on shorter baselines with smaller mixing amplitudes.

\end{abstract}

\flushleft
\end{titlepage}

\pagebreak

A principal feature of all of the experimentally known fundamental
fermions{\cite{PDG}} may be summed up in the following manner:  The
fermions may be grouped into three ``families'', conventionally
assigned by mass, consisting of two color triplets of quarks, one with
electric charge $+2/3$ and the other $-1/3$, one lepton with electric
charge $-1$, their antiparticles and a neutrino.  Each ``family'' of
these fermions fills out fifteen $(\half,0)$ representations of the
Lorentz group. Fourteen of these come in pairs with conjugate color and
(electric) charge quantum numbers so that they may be reconstructed
into seven Dirac bispinor representations. This is accomplished by
using charge conjugation under the Lorentz group\footnote{This should
not be confused with the larger CP operation used earlier in connection
with Majorana objects~\cite{Case+}.} to convert one member of each pair
into the requisite $(0,\half)$ representation{\cite{PR}}.  Despite this
construction, and the fact that these pairs are not conjugate in their
electroweak quantum numbers, the Lagrangian of the Standard Model (SM)
does not intrinsically violate conservation of the weak interaction
quantum numbers, courtesy of the chiral projections included in the
interactions.

The exceptional case is that of the neutrino, which has no known
partner representation. That such a partner should exist has long been
suggested{\cite{Bruno,BP}} and is especially evident in ``vector-like''
Grand Unified Theories (GUTs). A satisfactory explanation of the
stringent bounds on the mass of each of the three different flavors of
neutrinos has been developed in this context in terms of the so-called
``see-saw'' mechanism{\cite{GRS}}. This mechanism postulates that 
since the missing partner $(\half,0)$ representation carries no SM
quantum numbers at all, and so is ``sterile'' with respect to all SM
interactions, it may naturally acquire a large (GUT scale) Majorana
mass.  While the usual (active) $(\half,0)$ representation could also,
in principle, develop a Majorana mass, the associated scalar field must
carry weak isospin, $I_W = 1$, so that mass is usually assumed to be
zero. The effect of this is to suppress the induced Majorana mass of
the neutrinos active in the SM from the value common for Dirac fermion
masses in the SM by a factor of the ratio of such masses to something
very roughly on the order of the GUT-scale mass\footnote{Note that, in
the usual discussions of the ``see-saw'', the mass term that couples
the active $(\half,0)$ representation to itself as a Lorentz charge
conjugate $(0,\half)$ representation is referred to as $m_L$ while the
similar term for sterile neutrinos is referred to as $m_R$.  Although
that notation is natural under the assumption of Dirac neutrinos, here,
since the neutrinos under discussion are massive Weyl (Majorana), that
identification can lead to confusion.  Hence, we use ``active'' and
``sterile'' throughout this paper.}.

While the prospect of providing a ``natural'' explanation for the small
scale of active neutrino masses is pleasing, there is no principle
requiring that the sterile mass be large.  Here we discuss an equally
valid scenario based on a view of quark-lepton symmetry, which is
subject to a general experimental test that will soon be undertaken.

We start from the facts that the known Dirac fermion masses span a
range of almost six orders of magnitude and that those of the neutrinos
must be at least five to six orders of magnitude smaller still. We
allow for the possibility that the true origin of these masses is still
not understood and set aside the see-saw. We next recall that it is
charge-conservation, for various charges, which eliminates the
possibility of Majorana mass terms for each of the fourteen spinor
representations that make up the known Dirac bispinors.

Now we recall an old conjecture{\cite{PSD}}:  That there is indeed a
$(\half,0)$ representation for a sterile neutrino to form an eighth
pair with the known active neutrino for each generation (or family) of
fermions.  Recalling that all other individual fermion number (baryon
number, muon number, etc.) violations seem to be strongly suppressed,
we are led to examine the possibility that this is true for neutrinos
as well.  This leads naturally to the conclusion that, while the
sterile neutrino may have a Majorana mass, it should be expected to be
small compared to the Dirac mass available to the pair of neutrino
representations which can be formed into a Dirac bispinor.  We must
then simply accept the fact that the Dirac mass for this bispinor is,
itself, very small to satisfy experimental constraints, although there
are interactions which could exist that would modify the interpretation
of these constraints{\cite{clouds}}.

Many have conjectured{\cite{PS}} that there should be some parallel
(for example, right-chiral interaction) quantum number so that some
sort of neutrino number remains.  A related point has been made by
Cahill{\cite{Kevin}}, that a Majorana neutrino mass would violate
lepton number (L) conservation, and hence also the difference between
that and baryon number (B). The experimentally reported suppression of
proton decay{\cite{pdecay}} provides strong support for an assumption
that matrix elements for B-L violation are extremely tiny. Thus, we
consider it viable to investigate the implications of the point of view
that the Majorana mass terms of neutrinos are also quite small compared
to Dirac neutrino mass terms.

This very general scenario leads to a quite well-defined class of
predictions for neutrino properties and experiments. Denoting the Dirac
mass connecting the active and sterile neutrinos by $M$ and the
Majorana mass of the sterile neutrino by $m$, the conjecture that $m <<
M$ leads to the conclusion that the neutrino states are
pseudo-Dirac.\footnote{Here we use pseudo-Dirac to mean a pair of
Majorana neutrinos with masses so nearly degenerate that, for many
purposes, the linear combinations appropriate to a particle and an
antiparticle are, to a good approximation, eigenstates of the mass
matrix.  This, we believe, is the general usage{\cite{PSD}}.
Wolfenstein introduced the term{\cite{Wolf}} to refer to a particular
model in which the two $(\half,0)$ representations used to produce the
Dirac bispinor were both active under the weak $SU(2)$, but were
coupled to different charged leptons.} That is, the neutrino field
eigenstates will be a pair of almost degenerate states and a neutrino
will propagate almost as if it were a Dirac fermion. However, with a
long oscillation length, it will transform between the active and
sterile components, with almost {\em maximal} mixing (due to the almost
complete degeneracy of the resulting Majorana mass eigenstates). This
would be most simply expected to be true for each neutrino type
separately. (See also Ref.\cite{Kevin}.)

The immediate implication is that when maximal mixing is observed in
neutrino oscillations, it will be between active and sterile types.
Hence, the long baseline from the Sun leads to the conclusion that if
the signal diminution observed in solar neutrino
experiments{\cite{solar}} is due to vacuum neutrino oscillations, then
the oscillation that is occuring is from active electron neutrinos to
sterile (anti)neutrinos. It follows that the SNO experiment{\cite{SNO}}
is predicted to observe a reduction in the neutral current signal equal
to that already found in the charged current signal. This prediction
has also been made in Ref.\cite{Kevin}.

Similarly, in the observed atmospheric oscillations{\cite{SK}}, the
long baseline suggests that the oscillation is from muon to sterile
neutrinos. Although this is not favored by the current data set,
neither is it inconsistent at present. The scenario discussed here
predicts that additional data will find a diminishing signal for
active-active oscillation.

We should, however, note a possibility which is difficult to encompass
within unified models, but may nonetheless occur: The conjugate partner
to the active neutrino representation of one family may be the active
neutrino representation of a different family{\cite{Wolf}}. This would
appear to violate quark-lepton symmetry and leaves one uncertain about
whether or not there must be sterile partner representations. However,
this possibility matches more closely with the preferred interpretation
of the observed atmospheric oscillations{\cite{SK}}, if the pair of
families involved are those of the muon and tauon. Note that this still
implies that the SNO experiment{\cite{SNO}} would observe the same
reduction of the neutral current signal as of the charged current
signal because the solar neutrino oscillation would still necessarily
involve a sterile neutrino partner. This conjecture raises the question
of whether or not some vestige of quark-lepton symmetry obtains in the
form of two additional sterile neutrino representations that mix only
with each other, perhaps still forming a pseudo-Dirac bispinor.

We also note that there is the possibility of a modified ``see-saw'',
in which the sterile neutrino mass matrix in family space may have a
large scale but a rank less than three.  In this case one or two of the
families may have neutrinos that are poorly described as pseudo-Dirac
without affecting the remaining families.  For example, if the rank is
one, there are two zero eigenvalues of the $m$ matrix, were it
diagonalized by itself.  The embedding of that matrix in the larger
mass matrix for neutrinos can easily change two Dirac neutrinos into
pseudo-Dirac neutrinos, or could lead to Majorana neutrinos with masses
well-separated on the scale of the Dirac masses.

A priori, no definite predictions are made for flavor oscillations of
the type reported to be observed by the LSND
collaboration{\cite{LSND}}. However, as the Dirac mass terms are larger
than the Majorana mass terms in the particular scenario discussed here,
and the Dirac mass terms may be presumed (on the basis of quark-lepton
symmetry) to be analogous in structure to those found in the quark
sector, flavor mixing should be expected to occur with shorter
oscillation lengths and modest mixing amplitudes, i.e., much less than
maximal.  This is certainly consistent with the experimental
reports{\cite{LSND,KARMEN,CHOOZ}} to date.  Additional support for this
scenario of small flavor mixing between pseudo-Dirac neutrinos may be
found in recent discussions of possible interference effects modifying
the end point spectrum in Tritium beta decay{\cite{TRIT}}, when taken
in combination with existing limits{\cite{HM}} on Majorana neutrino
masses from neutrinoless double beta decay experiments.

Finally, we conclude that the strength of the neutral current signal of
the SNO experiment is crucial to determining the viability of any
pseudo-Dirac bispinor scenario.

We acknowledge stimulating conversations on this subject with Bill
Louis. This research is partially supported by the Department of Energy
under contract W-7405-ENG-36, by the National Science Foundation and by
the Australian Research Council.

\end{document}